# Life history shapes variation in egg composition in the blue tit *Cyanistes caeruleus*

Cristina-Maria Valcu[1], Richard A. Scheltema [2,3], Ralf M. Schweiggert[4], Mihai Valcu[1], Kim Teltscher[1], Dirk M. Walther[2], Reinhold Carle[4,5] & Bart Kempenaers [1]

Maternal investment directly shapes early developmental conditions and therefore has long-term fitness consequences for the offspring. In oviparous species prenatal maternal investment is fixed at the time of laying. To ensure the best survival chances for most of their offspring, females must equip their eggs with the resources required to perform well under various circumstances, yet the actual mechanisms remain unknown. Here we describe the blue tit egg albumen and yolk proteomes and evaluate their potential to mediate maternal effects. We show that variation in egg composition (proteins, lipids, carotenoids) primarily depends on laying order and female age. Egg proteomic profiles are mainly driven by laying order, and investment in the egg proteome is functionally biased among eggs. Our results suggest that maternal effects on egg composition result from both passive and active (partly compensatory) mechanisms, and that variation in egg composition creates diverse biochemical environments for embryonic development.

[1] Department of Behavioural Ecology and Evolutionary Genetics, Max Planck Institute for Ornithology, 82319 Seewiesen, Germany. [2] Department of Proteomics and Signal Transduction, Max Planck Institute of Biochemistry, 82152 Martinsried, Germany. [3] Biomolecular Mass Spectrometry and Proteomics, Bijvoet Center for Biomolecular Research and Utrecht Institute for Pharmaceutical Sciences, Utrecht University, 3584 CH Utrecht, The Netherlands. [4] Plant Foodstuff Technology and Analysis, Institute of Food Science and Biotechnology, University of Hohenheim, 70599 Stuttgart, Germany. [5] Biological Science Department, King Abdulaziz University, Jeddah, Saudi Arabia. Correspondence and requests for materials should be addressed to B.K. (email: b.kempenaers@orn.mpg.de)

Maternal effects occur when the maternal phenotype or environment influences offspring phenotype independently from the inherited genotype[1]. Early developmental stages are particularly sensitive to such effects, which can have long-lasting consequences for an individual's survival and reproductive performance[2,3]. Prenatal maternal investment is therefore a major component of the maternal effects with important consequences for both parent and offspring fitness[4].

In birds, egg quality (here used to indicate both egg size and composition) directly determines offspring viability and performance[5]. Bird eggs are not only equipped with the macro-nutrients, micro-nutrients and the complex biochemical machinery required for embryo development, but also with an effective defensive system designed to protect the embryo and the egg content from environmental and biological hazards[6]. Egg composition varies both between and within clutches, and may depend on, for example, temperature[7], storage time[7–11], level of competition[12], infection risks[13] or embryo sex[14]. Within genetic, physiological and environmentally defined limits, females can adjust both egg size[5] and composition[11–16] to meet specific requirements of the developing embryos, favour offspring with higher breeding potential or compensate for specific risks. When females can predict the environmental conditions their offspring will encounter, they may thus alter offspring development and phenotype and increase their fitness via phenotypic plasticity, potentially influencing the evolutionary response to environmental change[17].

Known mediators of maternal effects include egg hormones[18], carotenoids[19–21], defensive proteins (e.g., antibodies[22], lysozyme[10,23], avidin[23]) and calcium[24]. There are indications that other egg components such as vitamins[11,25,26], fatty acids[8,27] and other defensive proteins (e.g., ovotransferrin[9,28]) also mediate maternal effects, although a correlation between their abundance and offspring phenotype has not yet been shown. Defensive proteins are obvious candidates for mediating maternal effects, due to their important and well-understood protective function both pre and post hatching[10,29–31]. However, recent studies have shown that egg proteins cover a wide and largely unexplored functional range[32,33]; many of them could potentially mediate maternal effects. Although protein resources available during embryonic development clearly determine post-hatching development and phenotype[34], the concentrations of most egg proteins and how these concentrations vary among eggs remains unknown.

To assess whether and how birds differentially deposit egg proteins, we analysed the albumen and yolk proteome of 114 eggs from 39 clutches of free-living blue tits (*Cyanistes caeruleus*). Blue tits are small, socially monogamous passerines which produce only one clutch per year (replacement clutches can occur after brood failure). Females typically lay one egg per day (but longer laying gaps occur) until the clutch is complete (7–15 eggs). Total clutch mass often exceeds female body mass. Blue tits are income breeders, i.e. they acquire all nutrients for egg formation from their daily food[35]. Hence, variation in female investment is expected both between and within clutches. This is important, because earlier work suggested that most of the variation in embryonic development rate in this species is due to egg composition[36].

To explore how the egg proteome could mediate maternal effects in blue tits, we considered the most relevant life-history traits known to influence egg quality in birds, i.e., the age of the breeding female, the start of egg laying (lay date), clutch size, egg weight, the egg's position in the laying sequence (laying order), embryo sex (male vs. female) and paternity (within-pair: sired by the social mate vs. extra-pair: sired by another male in the population). We interpret our results in an ecological, life-history and phenological context, based on a long-term study of this blue tit population, as well as in the complex biochemical context of the egg composition. We show that clutch size, female age, laying order and paternity predict egg quality (weight and composition). The egg proteome and the yolk lipids and carotenoids form specific chemical signatures which are partly correlated with these life-history traits. The observed variation in the egg proteome strongly indicates that egg composition has a high potential to mediate maternal effects on offspring phenotype.

## Results

**Blue tit egg composition**. During the breeding season of 2014, we collected blue tits eggs from a population breeding in nest-boxes in a forest in south Germany (for details on the study design, see Methods). We assessed the quality of each egg in terms of weight and macro-nutrient composition (albumen and yolk total protein content, and yolk lipid content). Additionally, we used liquid chromatography-tandem mass spectrometry to measure yolk carotenoids (Supplementary Fig. 1).

Egg weight and macro-nutrient concentrations (Table 1) were similar to those previously reported in tits[8,37]. Egg composition was largely independent from egg weight with the exception of yolk lipids (Fig. 1). Thus, although heavier yolks contained more lipids (Supplementary Data 1), lipid concentration decreased with egg weight (Fig. 1). Total carotenoid concentrations (Table 1) were similar to those reported in rural great tits *Parus major*[11]. In agreement with earlier studies on tits[20,38], lutein, zeaxanthin and β-carotene were the most abundant carotenoids (Table 2). We detected three other minor carotenoids (β-zeacarotene, γ-carotene and a γ-carotene (Z)-isomer) which are known from plants, fungi and bacteria[39], but have not previously been found in avian eggs. Total carotenoid concentration did not depend on yolk weight (Fig. 1) or on yolk lipid concentration (Supplementary Data 1).

**Description of the blue tit egg proteome**. Using high-resolution mass spectrometry, we identified 171 albumen and 156 yolk protein groups (of which 149 and 139, respectively, were identified based on two or more peptides), representing 233 different protein functions: 161 in the albumen and 142 in the yolk (Supplementary Data 2). In comparison, in chicken eggs, 158 proteins have been identified in the albumen[32] and 119 in the yolk[33]. Using a label-free approach, we measured the abundance of 85% of the albumen and 86% of the yolk proteins (Supplementary Data 3). The estimated protein concentrations (see Methods) covered a dynamic range of six orders of magnitude, with the 10 most abundant proteins accounting for more than 90% of the egg's protein content (Supplementary Data 3, in bold). Most of the abundant blue tit egg proteins (ovalbumin-like, similar to ovalbumin-related protein Y, ovotransferrin, TENP-like, ovomucin/Mucin-5B, apolipoprotein B-100, vitellogenin-1-like, vitellogenin-2-like) are known as major egg proteins from domesticated bird species[32,33,40]. Others (alpha-2-macroglobulin-like protein 1-like, ovostatin-like) are minor proteins of the chicken egg white[41]. The proteins were classified into 21 functional categories according to the Munich Information Center for Protein Sequences (MIPS) Functional Catalogue (FunCat)[42] (Supplementary Data 2 and 4, Fig. 2).

**Explaining variation in egg weight and composition**. Egg quality varied among blue tit eggs both in terms of weight and composition (Table 1). Female identity explained about 79.7% of the variation in egg weight (intercept-only linear mixed-effect model with female identity as random factor). This is comparable to the ~70% between-clutch variation reported for blue tits[43] and other species[44]. In comparison, egg composition varied more

### Table 1 Summary of egg weight and composition

| Parameter | Egg weight (g) | Albumen weight (g) | Yolk weight (g) | Albumen protein | | Yolk protein | | Yolk lipids | | Total carotenoids | |
|---|---|---|---|---|---|---|---|---|---|---|---|
| | | | | (mg/egg) | (mg/g wet albumen) | (mg/egg) | (mg/g wet yolk) | (mg/egg) | (mg/g wet yolk) | (µg/egg) | (µg/g wet yolk) |
| Mean | 1.16 | 0.83 | 0.27 | 43.4 | 52.4 | 16.3 | 61.2 | 58.7 | 222.1 | 11.8 | 45.8 |
| Min | 0.85 | 0.60 | 0.19 | 28.0 | 42.1 | 11.5 | 42.4 | 47.5 | 170.7 | 1.8 | 7.0 |
| Max | 1.52 | 1.13 | 0.35 | 58.7 | 63.1 | 22.3 | 72.3 | 79.5 | 279.8 | 55.7 | 203.5 |
| SD | 0.11 | 0.09 | 0.03 | 6.0 | 4.1 | 2.2 | 5.4 | 6.6 | 21.8 | 9.3 | 36.7 |
| CV (%) | 10 | 11 | 11 | 14 | 8 | 14 | 9 | 11 | 10 | 79 | 80 |
| Between-clutch variation (%) | 80 | 81 | 72 | 60 | 13 | 51 | 18 | 20 | 9 | 0 | 4 |
| Within-clutch variation (%) | 20 | 19 | 28 | 40 | 87 | 49 | 82 | 80 | 91 | 100 | 96 |

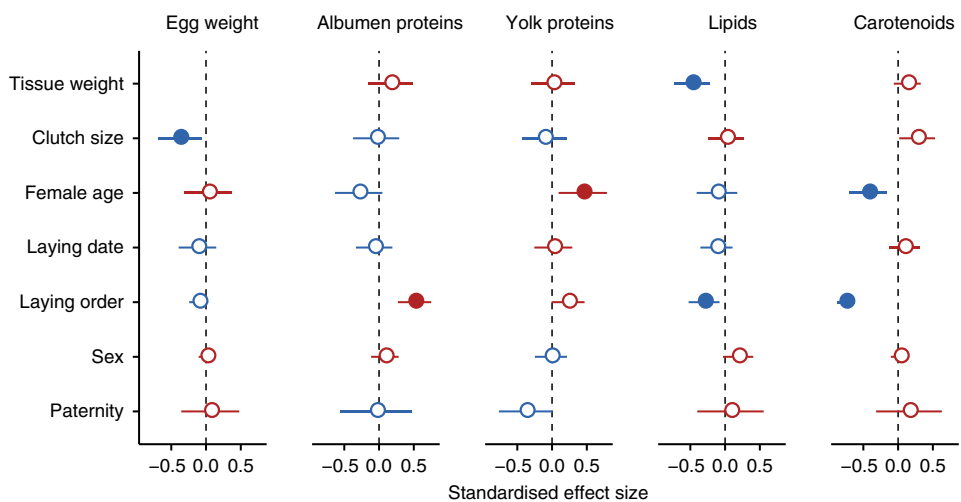

**Fig. 1** Predictors of blue tit egg weight and composition. Shown are standardised effect sizes and 95% confidence intervals (CIs) from linear mixed-effect models with female identity as random intercept and laying order as random slope. Dependent variables (total egg weight in mg, mg proteins, mg lipids or µg carotenoids/g wet sample) were standardised. Female age estimates are for adults relative to yearlings, embryo sex estimates are for males relative to females and paternity estimates are for extra-pair sired eggs relative to within-pair. Models with paternity as predictor are applied to first eggs only and control for the other predictors. Model estimates are in Supplementary Data 5 and 1. Negative estimates are in blue and positive estimates in red. Filled symbols indicate statistically significant differences (after false discovery rate correction). There were no significant interactions among the predictors of egg composition (all adjusted $p$ values of two-way interactions > 0.05)

### Table 2 Concentrations of the yolk carotenoids

| Carotenoid | Proportion of samples (%) | Mean concentration (µg/g yolk) | CV (%) |
|---|---|---|---|
| Lutein | 100 | 22.1 | 89 |
| Zeaxanthin | 100 | 7.2 | 87 |
| β-Carotene | 100 | 3.6 | 69 |
| β-Zeacarotene | 67 | 3.5 | 175 |
| γ-Carotene | 100 | 1.7 | 134 |
| γ-Carotene (Z)-isomer | 100 | 1.2 | 92 |
| Unidentified carotenoids | 100 | 7.7 | 91 |

within clutches (Table 1). Albumen and yolk weights varied isometrically with egg weight (Supplementary Data 5).

Between clutches, egg quality can vary due to differences in parental quality, environmental factors (e.g., habitat, temperature, nutrient availability, parasite infestation) or as a result of female reproductive decisions. For example, in passerines, between-clutch variation in egg quality is related to differences in clutch size[45,46], female age[46,47] and laying date[8,48]. Below we describe how these traits relate to egg quality in our blue tit population.

Egg production is energetically costly and the resources available to breeding females are typically limited[49,50]. Hence, life-history theory predicts that females laying larger clutches may compensate the costs by laying smaller eggs[51,52]. Indeed, clutch size was the only significant predictor of egg weight ($p = 0.041$, Supplementary Data 5), with a 2.5% decrease in the average egg weight for each additional egg in the clutch (Fig. 1). Egg composition did not vary with clutch size (Fig. 1).

Maternal investment often increases with female age[44], either because adult females occupy better territories, have better foraging skills or higher willingness to invest in the current breeding attempt due to lower prospects of future reproduction. In our study, neither clutch size (Supplementary Data 6) nor egg weight (Fig. 1) varied with female age. However, there was a significant interaction ($p = 0.016$) between female age and laying order on egg weight: in yearling females, egg weight increased with laying order, while in adult females it decreased (Supplementary Data 5). Yearling females laid lighter eggs than adult females at the beginning of the clutch (mean difference = 0.428, SE = 0.151, $z = 2.833$, $p = 0.022$), which may have allowed them

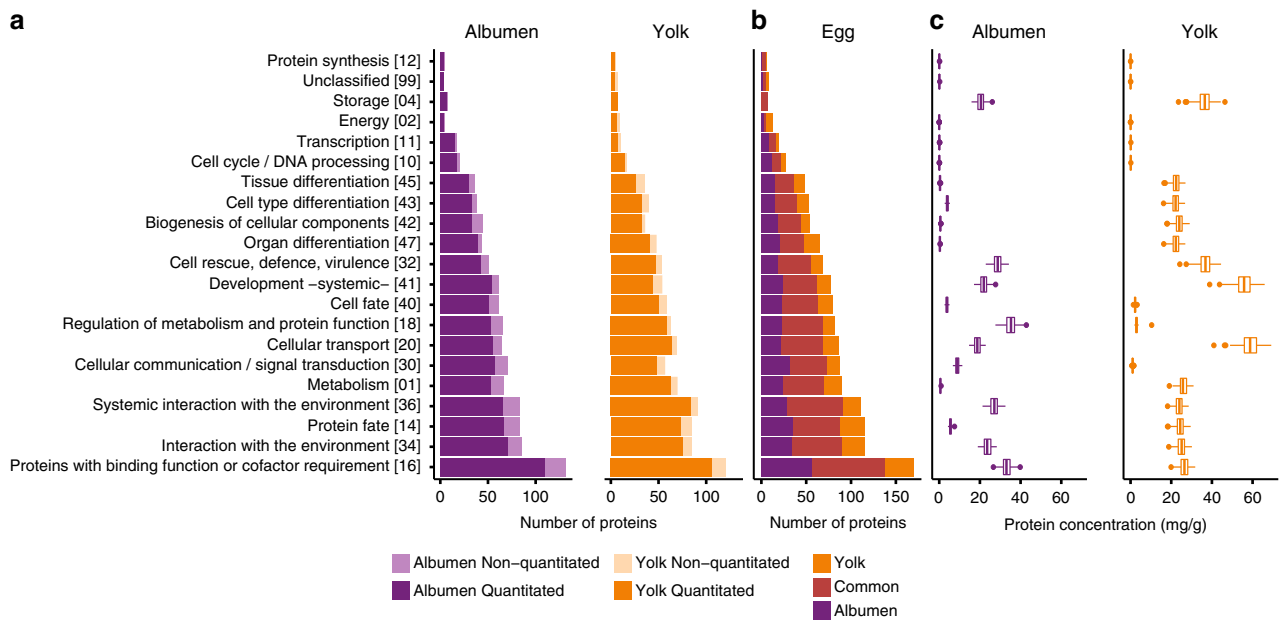

Fig. 2 Summary of the blue tit egg proteome. **a** Number of albumen and yolk proteins for different functional categories (based on the MIPS Functional Catalogue). FunCat identifier numbers are given between square brackets. Shown are frequencies for first-level FunCat terms (for second- and third-level FunCat terms see Supplementary Data 2). One protein may be represented under multiple terms (individual proteins had between 1 and 16 functions). Proteins that were not quantitated are shown in lighter shades. **b** Total number of egg proteins in each functional category. The number of proteins unique to yolk and albumen and those detected in both are indicated with different colours. **c** Protein concentrations combined for each functional category in the albumen and yolk. Shown are box plots with the median (vertical line), the 25th and 75th percentile (box), and the outliers (dots). Raw and summary data are given in Supplementary Data 3 and 4, respectively

to start laying as early in the season as adult females[53]. Eggs of adult females contained yolks with higher protein concentrations than those of yearling females (Fig. 1). Opposite to a previous report on blue tits[12], eggs of adult females contained lower yolk carotenoid concentrations than those of yearling females (Fig. 1).

In many species, breeding females are under selective pressure to start laying early in the season because reproductive success typically declines over the season[54]. In great tits[46] and in this blue tit population (Supplementary Data 6), adult females typically start laying earlier in the season than yearling females. However, in the 2014 breeding season, there was no temporal segregation of egg laying between yearling and adult females (Supplementary Note 1). Previous studies showed that egg weight increased with laying date in tits, possibly as a consequence of increasing food availability as the season advanced or because of the decreasing metabolic cost of thermoregulation due to the rise in temperature[48,49]. Yolk carotenoid levels in blue tit eggs can also increase with laying date, likely reflecting an increase in the availability of carotenoid-rich food[12,55]. However, in this study, neither egg weight nor egg composition significantly varied with laying date (Fig. 1), which may be specific to the short and synchronous 2014 breeding season.

Within a clutch, egg quality may vary due to temporal variation in food availability, in female condition or in the metabolic costs for individual maintenance (e.g., due to thermoregulation), or because of female reproductive decisions. Females can for example alter egg quality to favour certain offspring[14], to protect early-laid eggs from the risks of prolonged storage (e.g., dehydration[7], infection[9–11]) or to adjust the egg's developmental rate to achieve more synchronous hatching[36,56]. Below we describe the variation in blue tit egg weight and composition in relation to laying sequence, embryo sex and paternity.

Laying order is particularly relevant in species with large clutches, where early-laid eggs may be in the nest for up to 2 weeks before the onset of incubation. Previous studies in blue tits have found that egg size either increased with laying order[57] or decreased at the end of the clutch[58]. In our population, laying order was the strongest predictor of egg quality. Egg weight increased with laying order in clutches laid by yearling females (0.006 g or 0.5% per egg) and decreased in those of adult females (0.006 g or 0.5% per egg) (Supplementary Data 5), suggesting that, at least in some populations/seasons, yearling and adult females take different within-clutch investment "decisions" or experience different physiological or nutritional constraints. For example, adult females may experience physiological exhaustion, while yearling females may benefit more from increasing resource availability and decreasing maintenance costs as the season advances[48,49]. Laying order was also the main driver of egg composition. Lipid and carotenoid concentrations decreased with laying order, as shown previously in tits[8,11], while protein concentrations increased (significantly in the albumen and weakly in the yolk, Fig. 1, Supplementary Data 1). The decrease of carotenoid levels with laying order (Fig. 1) was independent of the yolk's lipid content (Supplementary Data 1) and was observed in all measured carotenoids (Fig. 3a).

Previous studies suggested that female blue tits adjust their primary sex ratio depending on clutch size[59], female age[59], mate quality[59–61] or the timing of laying[59], and that offspring sex depends on laying order[57]. Although no sexual dimorphism in egg size has been found in blue tits, egg size can vary with laying order differently for male and female eggs[57]. In our study population, sex ratio did not vary systematically with female age, laying date and clutch size, and embryo sex was independent of laying order (Supplementary Data 7). Furthermore, we found no significant sex bias in egg weight or composition (Fig. 1).

In blue tits and other species, extra-pair offspring can outperform their within-pair half-siblings[62], an effect believed

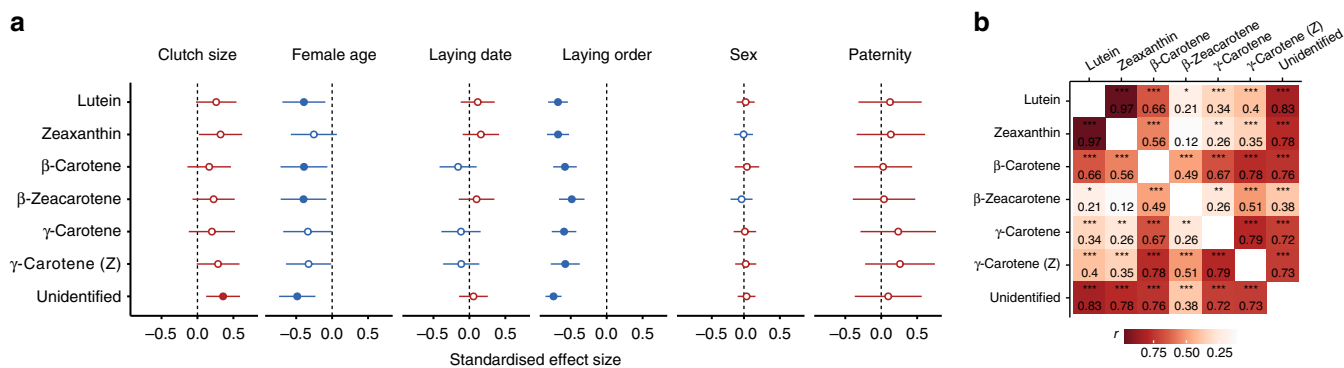

**Fig. 3** Blue tit egg yolk carotenoids. **a** Correlation matrix heat map (Pearson's *r*-values). Affinity propagation clustering of yolk carotenoids returned a single cluster. N = 109 eggs from 38 clutches; *$p < 0.05$, **$p < 0.01$, ***$p < 0.001$. **b** Predictors of the concentration of individual yolk carotenoids. Shown are the standardised effect sizes and 95% confidence intervals obtained from linear mixed-effect models with each carotenoid as dependent variable and with female identity and laying order as random intercept and slope, respectively. Each graph shows the effect of a predictor on the concentrations of individual carotenoids, with negative estimates in blue and positive ones in red. Female age estimates are for adults relative to yearlings, embryo sex estimates are for males relative to females and paternity estimates are for extra-pair sired eggs relative to within-pair. Models with paternity as predictor are applied to first eggs only and control for the other predictors. Model estimates are in Supplementary Data 10

to reflect maternal effects mediated by laying order[63]. Extra-pair eggs are mostly laid early in the laying sequence, presumably due to a decline in the frequency of extra-pair copulations after the start of laying[64]. As a consequence, extra-pair eggs also hatch earlier than within-pair eggs, which seems to explain most of the measured differences between within- and extra-pair young[63]. Our results support these findings, i.e., we only found extra-pair eggs among the first eggs of the sampled clutches (Fisher's exact test, $p < 0.001$). The occurrence of extra-pair paternity (between clutches) was not predicted by laying date, clutch size or female age (Supplementary Data 7). It has been reported that the sex ratio of extra-pair young is biased towards males in blue tits[62] (but see ref.[65]); however, we found no difference in sex ratio between within-pair and extra-pair eggs (Fisher's exact test, $p = 0.31$). Extra-pair and within-pair sired first-laid eggs were similar in weight and composition (Fig. 1).

**Explaining variation in the egg proteome.** We investigated variation in the blue tit egg proteome both at the level of individual proteins (Fig. 4) and at the level of functional categories (Fig. 5). As for the overall egg composition, laying order was the most important predictor of proteome variation.

In both egg compartments, the effect of laying order varied widely among individual proteins, similarly to previous reports[28]. The concentrations of 35 albumen and 23 yolk proteins increased, while the concentrations of 26 albumen and 12 yolk proteins decreased significantly with laying order (Fig. 4). Of the 10 most abundant egg proteins, 9 increased in concentration with laying order, which explains the overall increase in protein concentration, in particular in the albumen (Fig. 1).

The laying order effect was functionally biased. The combined concentrations of 8 functional categories of albumen proteins and 7 functional categories of yolk proteins were higher in late-laid eggs, while 6 functional categories of albumen proteins had higher concentrations in early-laid eggs (Fig. 5a). Protein categories involved in storage, metabolism and protein function regulation, and defence (FunCat 04, 18 and 32) increased in concentration with laying order in both egg compartments. Although early-laid eggs had lower overall protein concentrations than late-laid eggs, they contained higher relative abundances of proteins involved in metabolism, protein fate, biogenesis of cellular components, tissue differentiation and organ differentiation (FunCat 01, 14, 42, 45 and 47) in both egg compartments

(Fig. 5b). Because the deposition of egg constituents is subject to genetic, physiological and nutritional constraints, the higher deposition of particular proteins may require a compensatory reduction of the abundance of other proteins. The relative abundance of an egg protein (proportion of the total protein) reflects the degree to which this protein is preferentially deposited in the egg, at the expense of other proteins. Overall, 8 albumen and 10 yolk functional categories had higher relative abundances in early-laid eggs, while 6 functional categories of yolk proteins had higher relative abundances in late-laid eggs (Fig. 5b).

We also found a paternity effect on the egg yolk proteome (Fig. 4). Five proteins had significantly higher concentrations in within-pair eggs: apolipoprotein A-I, apolipoprotein B-100, fetuin-B, extracellular fatty acid-binding protein-like and a protein similar to complement C3. Of the 20 FunCat terms, 19 were more concentrated in within-pair eggs (14 statistically significant, Fig. 5a). However, these differences are mostly due to two of the three most abundant yolk proteins (apolipoprotein B-100 and vitellogenin-1-like), which had higher concentrations in within-pair eggs. Relative protein abundances did not differ among within- and extra-pair eggs (Fig. 5b). Note that these are between-clutch comparisons of first-laid eggs, because extra-pair paternity was only found in first-laid eggs.

Female age also had an effect on the yolk proteins, with higher concentrations of proteins involved in cellular communication/ signal transduction in adults (FunCat term 30, Fig. 5a). However, most functional categories of yolk proteins were more concentrated in the eggs of adult females, and the effect size on FunCat term 30 was among the lowest measured (Supplementary Data 8). Relative protein abundances did not differ among eggs of yearling and adult females (Fig. 5b). Thus, adult females appear to have an overall increased capacity to deposit yolk proteins compared to yearling females.

Although published data regarding the variation of individual egg proteins are sparse and restricted to a few defence and immunity-related proteins, our results generally support previous reports from other passerine species (Supplementary Note 2). Apart from laying order and paternity, the tested predictors did not explain variation in the concentration of individual egg proteins, with the following exceptions (Supplementary Data 3). (1) Eggs laid by adult females had higher concentrations of ceruloplasmin and of procollagen-lysine 2-oxoglutarate 5-dioxygenase 1 in the yolk compared to those laid by yearling

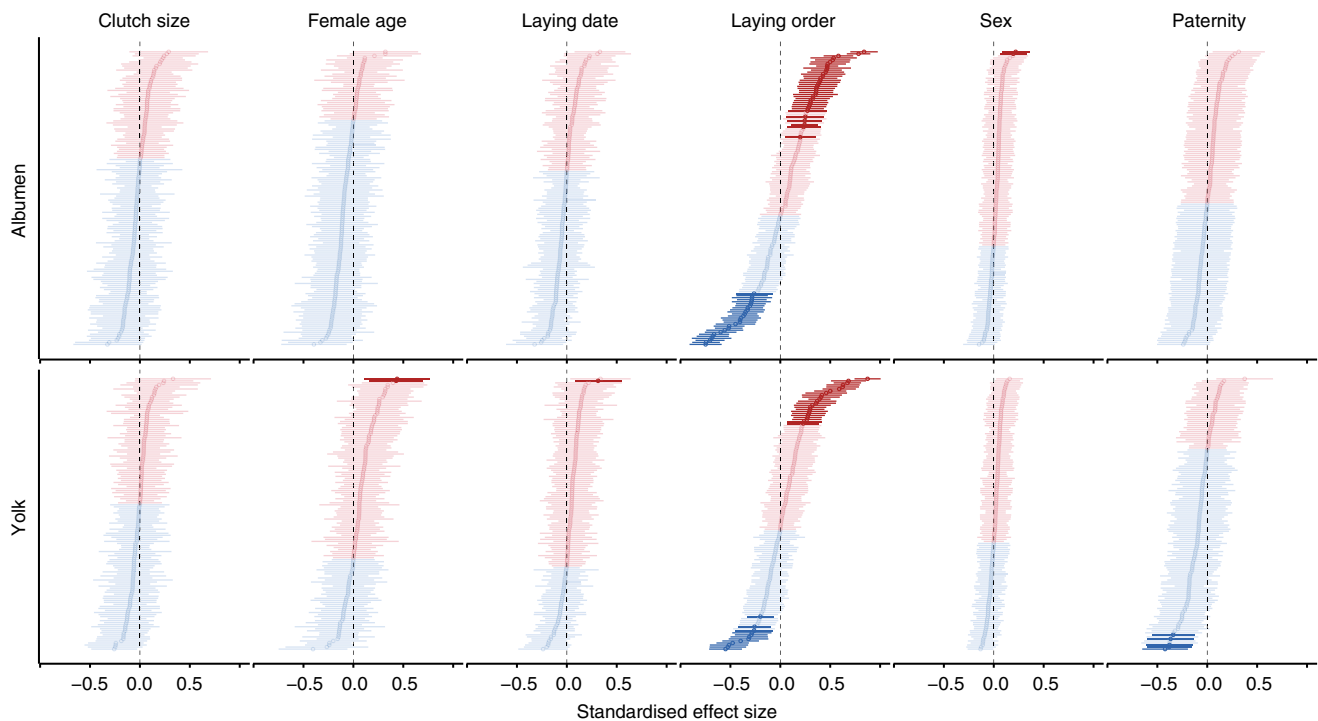

**Fig. 4** Predictors of variation in blue tit egg protein concentrations. Shown are standardised effect sizes and 95% confidence intervals from linear mixed-effect models with each protein in each egg compartment as a dependent variable and with female identity and laying order as random intercept and random slope, respectively. Each graph shows the effect of a predictor on the concentrations of individual proteins, with estimates shown in descending order. Female age estimates are for adults relative to yearlings, embryo sex estimates are for males relative to females and paternity estimates are for extra-pair sired eggs relative to within-pair. Models with paternity as predictor are applied to first eggs only and control for the other predictors. Model estimates are in Supplementary Data 3. Negative estimates are in blue and positive estimates in red. Darker shades indicate statistically significant differences (after false discovery rate correction). There were no significant interactions between predictors (all adjusted $p$ values of two-way interactions > 0.05)

females. (2) Eggs laid late in the season had higher concentrations of vanin-1 in the yolk than those laid early in the season. (3) Male eggs contained higher concentrations of cell growth regulator with EF hand domain protein 1 and of a protein similar to spore coat protein SP87 in the albumen than female eggs. Without further evidence and given no obvious functional relevance, we suggest interpreting these results with caution.

**Multivariate patterns of variation in egg composition**. Phenotypes are rarely an expression of single factors; they rather arise from complex networks of molecular interactions. Altering the offspring's phenotype through maternal effects on egg composition may therefore require the coordinated regulation of many different components. If so, selection may favour the covariation (positive or negative) of egg components which interact to determine the offspring's phenotype (the adaptive covariation hypothesis)[66–68]. Indeed, egg components with synergistic[69] or opposite[66] effects on offspring phenotype covary in passerine eggs. Theoretical work has also shown that, similar to quantitative genetics, maternal effects need be studied in a multivariate context[70,71].

To identify patterns of covariation among egg components, we applied an unsupervised machine-learning approach. We used affinity propagation clustering to uncover intrinsic patterns of protein covariation independent of any a priori chosen predictors. The affinity propagation algorithm partitioned the albumen and yolk proteomes into 7 and 9 groups of co-varying proteins, respectively, with several of these groups carrying functional signatures (Fig. 6a, c). The combined concentration of proteins in most clusters was predicted by one or more life-history traits, e.g. laying order (A2, A3, A5-7, Y4-6, Y9), female age (Y5), embryo sex (Y6) or paternity (Y5-8) (Fig. 6b, Supplementary Data 9). Pairs of protein clusters predicted by the same predictor(s) were typically correlated (Supplementary Fig. 2).

Egg proteins (individual proteins, functional categories and protein clusters) also covaried with yolk carotenoid concentrations (Supplementary Data 3 and 8, Supplementary Fig. 2), which was likely a consequence of variation with laying order. Functional categories of proteins related to defence and immunity were negatively correlated with carotenoid concentrations, both for albumen (FunCat 32, 34 and 36: defence, interaction with the environment and systemic interaction with the environment) and for yolk (FunCat 32: defence) (Supplementary Data 8). We found limited evidence for covariation between egg proteins and yolk lipids (Supplementary Data 3 and 8, Supplementary Fig. 2).

The carotenoids themselves formed a homogenous group (affinity propagation clustering produced a single cluster), with individual carotenoid concentrations closely following the variation of total carotenoid concentration (Fig. 3a, Supplementary Data 10). The concentrations of the most abundant carotenoids, lutein and zeaxanthin, were highly correlated and also strongly correlated with total carotenoid content. Correlations with β-carotene were somewhat weaker and those with the minor yolk carotenoids even smaller (Fig. 3b). A similar correlation pattern has been described for great tit eggs[25] and is likely explained by the tits' diet during the breeding season, largely consisting of caterpillars and spiders[72].

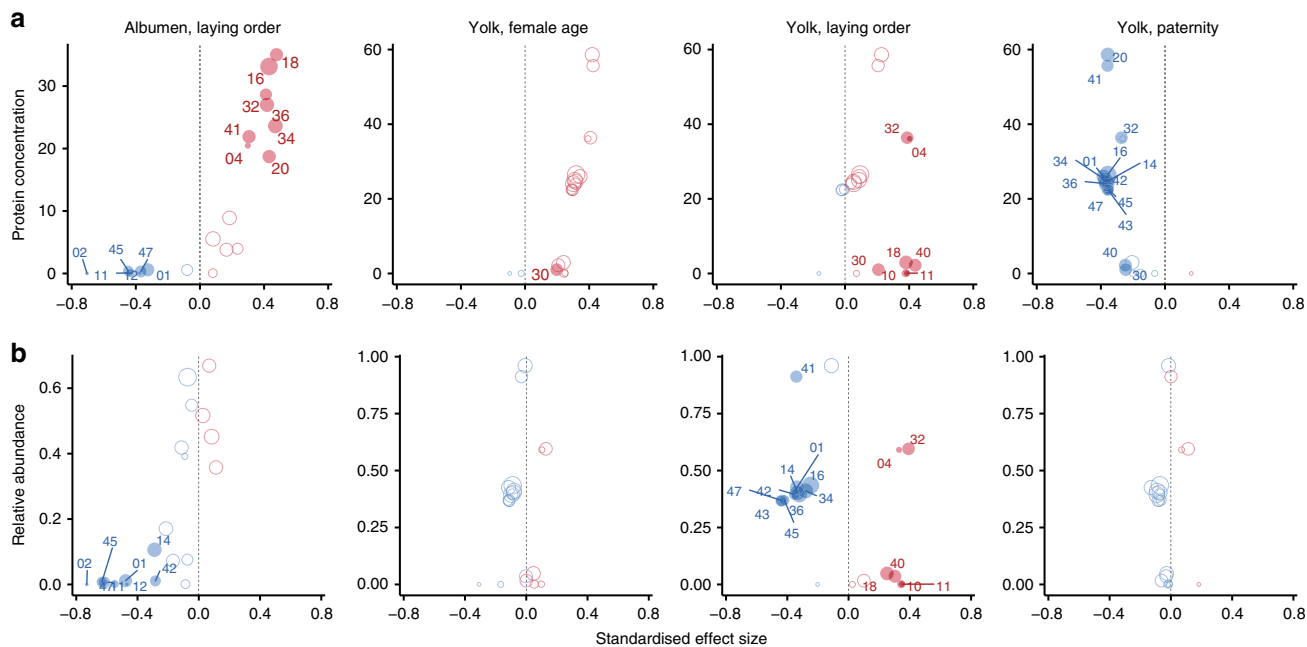

**Fig. 5** Predictors of the concentration and relative abundance of the functional classes of proteins. Shown are the standardised effect sizes for the significant predictors of the concentration and relative abundance (proportion of the total protein content) of the functional classes of proteins. Effect sizes are plotted against **a** the combined concentrations or **b** the combined relative abundance of all proteins belonging to a particular functional category. Negative estimates are in blue and positive estimates in red. Filled symbols mark significant estimates (after false discovery rate correction) with the numbers indicating the functional category identity (Supplementary Data 4). Symbol size is proportional to the number of proteins within each functional category. Model estimates are in Supplementary Data 8. There were no significant interactions between predictors (all adjusted $p$ values of two-way interactions > 0.05)

We also used affinity propagation clustering to identify homogenous subgroups of eggs based on their complete chemical composition. Input variables were the albumen and yolk protein clusters, yolk lipid and carotenoid concentrations as well as egg weight. The unsupervised algorithm produced three egg clusters (Fig. 6e). One egg cluster (E1) had higher lipid and carotenoid concentrations as well as a characteristic proteomic profile, while all egg clusters differed in their proteomic profiles (Fig. 6d, Supplementary Data 11). To compare these results with those obtained using the model-based approach, we mapped the six predictors on each egg cluster (Fig. 6f). The three egg clusters could be differentiated based on laying date, laying order, female age and paternity (Fig. 6g), but only laying order was statistically significant after false discovery rate correction ($p < 0.001$, Supplementary Data 12).

## Discussion

Egg quality is a highly variable life-history trait[3,44] with important consequences for offspring fitness[5]. Our results show that, in blue tits, female age and laying order drive variation of both components of egg quality, i.e., egg weight and composition (Figs. 1, 5). In addition, egg weight varies with clutch size (Fig. 1), while egg composition varies between within- and extra-pair sired eggs (Figs. 4, 5). Since egg weight is highly heritable[44], variation in egg weight may allow local adaptation. Meanwhile, variation in egg composition could facilitate offspring phenotypic variation through maternal effects. Overall, laying order was the strongest predictor of egg quality. Maternal effects mediated by laying order are important, particularly in species with large clutches, such as blue tits. Variation in egg quality with laying order might support differences in embryonic developmental rate[36], offspring growth rate[73], resistance to infections[10,28] and antioxidative protection[11,74], as well as mediate maternal effects on offspring sex[26] and paternity[63]. Below we discuss whether and how our results support these hypotheses and fit in with earlier work. We also examine the significance of egg composition variation for offspring phenotype and infer the mechanisms underlying the observed variation.

The yolk lipid concentration declined with laying order and egg weight (Fig. 1), while total egg protein concentration increased with laying order and did not vary with egg weight (Supplementary Data 1). In other blue tit populations, lipids also declined with laying order, but protein levels remained constant[8]. Other studies also showed that in breeding females of blue tits and other small income breeders, lipid depletion occurs at higher rates than protein depletion after egg laying[35,75–77]. This suggests that different constraints act on the biosynthesis and/or acquisition of these nutrients, and that the deposition of egg proteins is easier to sustain for females than that of lipids. The observed variation in egg nutrient concentrations has direct consequences for the offspring phenotype. Egg proteins serve mostly as structural and functional components, while lipids provide most of the energy required for embryo development and are essential as structural material for cell membrane synthesis. As a consequence, offspring hatched from protein-rich eggs are comparatively larger, while offspring hatched from lipid-rich eggs have larger nutrient reserves[78]. Blue tits are altricial birds with intense sibling competition. This means that hatchling size is more important than the amount of nutrient reserves, because it directly relates to competitive ability and hence facilitates access to food[79]. Given that the protein-to-lipid ratio increased overall with egg weight and laying order (Supplementary Data 1), offspring hatched from larger and late-laid eggs may therefore be "equipped" to be more competitive. A higher lipid concentration in early-laid eggs may also help to delay hatching until the incubation of the late-laid eggs is complete, such that hatching becomes more synchronous[56]. Concurrently, the higher protein content of late-laid eggs could support higher rates of embryonic growth, and thus reduce the effects of hatching asynchrony resulting from partial incubation before clutch completion[36]. The variation in egg composition with laying order may thus reflect both constraints on egg formation and female investment tactics related to avoiding asynchronous hatching. Note that in this study egg weight varied differently with laying order in clutches laid by yearling (inexperienced) and adult (experienced) females (Supplementary Data 5).

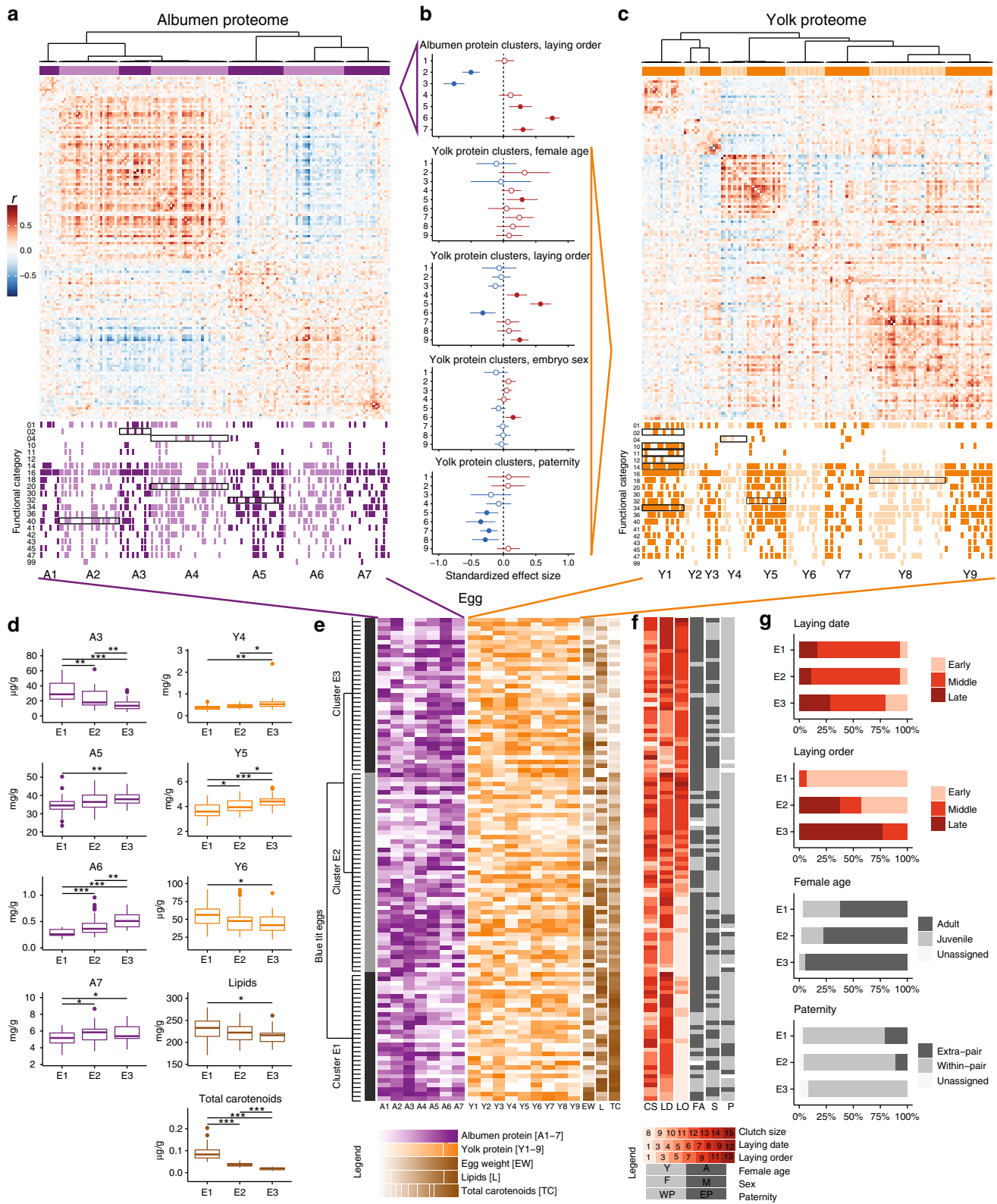

The concentration of yolk carotenoids declined with laying order (Figs. 1, 3), similarly to what has been reported in great

**Fig. 6** Affinity propagation clustering of blue tit eggs based on their composition. **a**, **c** Affinity propagation clustering and heat map showing between-protein correlation coefficients (Pearson's *r*) for the albumen (**a**) and yolk (**c**) proteins. Functional categories (first-level FunCat terms, see Supplementary Data 4) corresponding to each protein are mapped below the heat map indicating the 7 albumen and 9 yolk protein clusters. Black rectangles indicate significant enrichment of the protein clusters in each functional category as determined by Monte-Carlo simulation (see Methods). **b** Predictors of the albumen or yolk cluster combined protein concentrations. Only significant predictors are shown. Negative estimates are in blue and positive estimates in red. Filled symbols mark significant estimates (after false discovery rate correction). Female age estimates are relative to yearling females, embryo sex estimates are for males relative to females and paternity estimates are relative to within-pair. Model estimates are in Supplementary Data 9. **d** Description of the chemical composition of the three egg clusters. Shown are only egg components with significantly different concentrations in the three egg clusters (Supplementary Data 11). **e** Affinity propagation clustering of blue tit eggs based on their composition. A1-7 concentration of albumen proteins combined for each cluster (mg/g wet albumen), Y1-9 concentration of yolk proteins combined for each cluster (mg/g wet yolk), EW: egg weight (g), L: lipid concentration (mg/g wet yolk), TC: total carotenoid concentration (μg/g wet yolk), CS: clutch size, LD: laying date, LO: laying order, FA: female age (adult vs. yearling), S: embryo sex (male vs. female), P: paternity (extra-pair vs. within-pair). Affinity propagation clustering returned three egg clusters: E1, E2 and E3. **f** Life-history predictors mapped on the egg cluster. **g** Predictors of the egg clusters. Shown are the life-history variables which predict whether individual eggs belong to given clusters, based on functional chi-square tests (Supplementary Data 12). Laying order was the only significant predictor after false discovery rate correction. Non-binary variables were categorised into three levels based on equal intervals (see Methods)

tits[11], indicating diet-related constraints on the deposition of carotenoids. Carotenoids are liposoluble antioxidants obtained solely from the diet which increase offspring immunity and supply vitamin A to the developing embryo (Supplementary Note 3). Early-laid eggs are exposed to increased infection risks due to the longer storage period before incubation onset, and thus may benefit from the higher carotenoid concentrations and the increased immunity they impart[20,21].

The functional bias in the variation of protein concentrations with laying order suggests that egg proteomic profiles may also promote differences in the phenotypes of offspring originating from early- vs. late-laid eggs. Early-laid eggs had higher relative abundances of proteins involved in metabolism, protein fate, biogenesis of cellular components and tissue and organ differentiation (Fig. 5b), which may allow these offspring to hatch in a more developed stage than their siblings. Late-laid eggs, however, showed higher concentrations of proteins with storage, regulatory and defensive functions (in particular in the albumen) (Fig. 5a), possibly compensating for their lower carotenoid concentrations and providing higher protein resources for offspring growth. The negative correlation between carotenoids and defence- and immunity-related proteins (Supplementary Data 8) supports a compensatory investment into defence- and immunity-related proteins in late-laid eggs.

Paternity was another predictor of egg proteomic profiles. In blue tits, laying order and the associated hatching asynchrony[63] can explain the higher fitness of extra-pair young compared to their within-pair half-sibs[62]. Our results show that variation in egg composition with laying order can also contribute to these differences. Because extra-pair eggs are laid early in the clutch, they benefit from relatively higher lipid and carotenoid concentrations as well as from the functional proteomic profiles characteristic of such eggs. However, among first-laid eggs, four protein clusters had higher concentrations in the yolks of within-pair sired eggs (Fig. 6b), two of which were enriched in proteins involved in defence (FunCat 32, cluster Y5) and metabolism and protein function regulation (FunCat 18, cluster Y8). This potentially challenges the assumption that females favour extra-pair over within-pair offspring[80]; however, further work is needed to confirm this effect, elucidate the mechanisms and understand the consequences.

Given the egg complexity, it is reasonable to expect that different constraints or costs are associated with different egg components. For instance, egg lipids[8] and antioxidants[11,26] are typically physiologically and nutritionally limited, while the constraints on the deposition of egg hormones[68] and proteins[8,9] appear to be lower. In the context of parent–offspring conflict, one may speculate that within-clutch variation in the limited/costly egg components is predominantly driven by maternal interests, given the limited capacity of the female to deposit them. Conversely, for the less costly or less constrained components, variation in deposited amounts may be based on benefits to offspring. Females could compensate nutritional constraints on the deposition of some egg components through increased deposition of other, less costly components. Such a compensatory effect has been observed in eggs of gulls, where diet-derived carotenoids and vitamin E concentrations correlated inversely with testosterone concentrations[68,81].

While it is well established that egg composition varies within clutches of many species[7,9,10,28,74], it is still debated whether this variation is a passive consequence of changes in maternal diet and/or of the biochemical and physiological mechanisms of egg formation, or an active investment decision requiring specific transfer mechanisms[26]. Both passive and active mechanisms exist, and both can generate adaptive variation, but active mechanisms allow a higher level of maternal control[3]. Although these mechanisms cannot be distinguished solely from information on egg composition, the within-clutch variation of different egg components can provide some useful insights. Temporal variation in food quality and availability could parsimoniously explain both the decrease of yolk lipids and carotenoids and the increase of albumen proteins with laying order (Fig. 1), suggesting that the deposition of egg nutrients may require only passive mechanisms. However, protein variation with laying order was functionally biased in both the albumen and the yolk (Fig. 5), which likely required a selective regulation of protein synthesis, secretion and transfer. The negative covariation between carotenoids and defence- and immunity-related proteins (Supplementary Data 8) further suggests that the deposition of at least some egg proteins is compensatory. Previous research suggests that passive and active mechanisms are not mutually exclusive[3]. Our results are thus consistent with the hypothesis that egg nutrients are deposited through a combination of passive and active mechanisms.

Finding clusters of co-varying proteins in the egg proteome (Fig. 6a, c) is not unexpected: egg proteins can covary, e.g., because they are produced by the same cells, because they share transcription factors or because their synthesis is regulated by the same hormones[3]. Noteworthy however, the variation of different protein clusters was driven by different life-history predictors (Fig. 6b), indicating that the deposition of different groups of egg proteins can be regulated independently. If proteins are easier to deposit than other egg components, as this and other studies[8,35,75–77] suggest, the egg proteome could offer broad possibilities for adjusting egg composition. Furthermore, if variation in the concentration of individual egg proteins can support variation in offspring phenotype[10,22,23], the entire egg proteome could hold an enormous potential to mediate maternal effects. This appears to be a realistic scenario: the three identified egg clusters had different proteomic profiles (Fig. 6d) and were predicted by several of the investigated life-history traits (Fig. 6g).

Although this study considered some of the most relevant life-history traits of blue tits, these traits did not fully explain variation in the egg proteome, indicating that additional, unaccounted for factors may drive variation in egg composition (Fig. 6, Supplementary Fig. 2). An exhaustive screening of all potential factors influencing egg composition is virtually impossible. Moreover, the analysis of the egg's composition excludes measuring offspring traits (or at the very least alters offspring phenotype[82–84]). Thus, a complete description of the phenotypic consequences of egg composition variation remains inaccessible. Also, rather than being linked to measurable environmental, parental or offspring variables, part of the variation in egg composition may represent a diversified female bet-hedging strategy[85], aiming at equipping different eggs of the clutch to perform optimally under different and partly unpredictable environmental conditions[3]. The actual effect of egg composition on offspring fitness depends on the environment[81], conferring additional plasticity to the maternal effects. Identifying such a context-dependent maternal effect and determining its adaptive value is no trivial task. However, theoretical studies show that "multi-variate maternal effects can provide a clear signature of the past selective environment experienced by organisms"[70] even in the absence of information regarding their actual causes or fitness consequences. From this perspective, the egg proteome can open a new window into maternal effects on egg composition. The plasticity of the egg proteome indicates its high potential to mediate maternal effects while the covariation between proteins and other egg components reveals some of the proximate mechanisms underlying phenotypic plasticity in offspring development or probability of survival.

## Methods

**General field and laboratory methods**. We studied a blue tit population breeding in nest-boxes in an unmanaged, mixed-deciduous forest at Westerholz, southern Germany (48°08′ 26′′N 10°53′ 29′′E). The site hosts 277 nest-boxes and the population has been studied since 2007. We checked the nest-boxes weekly prior to and during nest building and daily close to the start of laying and during laying. During the 2014 breeding season, we collected eggs within a few hours after they had been laid and replaced them with dummy eggs. We collected the 1st and 9th egg from 39 of the 79 first clutches laid in this season (every second clutch in the population, Supplementary Fig. 3, Supplementary Data 13). From 10 of these clutches, we also collected every second egg laid. Of the total of 114 collected eggs, at most 2 were preceded by a laying gap (Supplementary Fig. 3). All nests were first breeding attempts and all breeding pairs except two were newly formed. Collected eggs were brought to the laboratory, kept at 12 °C until the next day, and then incubated at 38 °C for 24 h.

We measured egg and yolk weight to the nearest 0.01 mg with a Sartorius CP225D balance. We opened the eggs, removed embryo cells from the germinal disk area and stored them in 70% ethanol for subsequent DNA extraction. We collected the yolk and albumen and stored them at −80 °C until further processing. We calculated albumen weight by subtracting yolk and eggshell weight from the total egg weight.

**Ethics statement**. The study was carried out in accordance with the German Animal Welfare Act (Deutsches Tierschutzgesetz). All procedures were conducted under license from the Regierung von Oberbayern, Natur- und Artenschutzrecht (55.1-8676-TS-4-2014). Egg removal did not disturb breeding individuals: hatching success (% of eggs hatched) and fledging success (% of hatchlings that left the nest) averaged 87% ± 11 and 90% ± 14 for experimental nests compared to 85% ± 16 and 89% ± 15 for control nests.

**Genotyping and parentage analysis**. Samples were centrifuged at 8000 rpm for 1 min and the ethanol gently poured off. Remaining ethanol was allowed to evaporate at room temperature for 5–10 min. DNA was extracted using the QIAmp DNA Micro Kit from Qiagen following the protocol for Forensic Case Work samples. For the final step, DNA was eluted twice with 20 µl AE solution.

We genotyped all samples at 14 polymorphic microsatellite markers of which one was sex-chromosome linked: Pca3, Pca4, Pca7, Pca8, Pca9[86], POCC1, POCC6[87], Mcyµ4[88], ClkpolyQ[89], ADCbm[90], Phtr3[91], PatMP2-43[92], PK11, PK12 (Tanner et al., unpublished; EMBL Accession numbers: AF041465 and AF041466) and at the sex-chromosome linked marker P2/P8[93]. Microsatellite amplifications were performed in multiplexed PCRs using the Qiagen Type-it Microsatellite PCR Kit (Qiagen, Hilden, Germany) and primer mixes containing two to five primer pairs. The forward primer of each pair was fluorescently labelled with 6-FAM, VIC, PET or NED (Dye Set G5; Applied Biosystems, Darmstadt, Germany). Differences in amplification efficiency and dye strength of the primers were accommodated by adapting the primer concentrations in these mixes. Each 10 µl multiplex PCR contained 20–80 ng DNA (in max. 2 µl), 5 µl of the 2× Type-it Microsatellite PCR Master Mix and 1 µl of one of a primer mix. Cycling conditions were as follows. Mix 1: 5 min of initial denaturation at 95 °C; 15 cycles of 30 s at 94 °C, 90 s at 60 °C minus 0.3 °C per cycle, 60 s at 72 °C; 12 cycles of 30 s at 94 °C, 90 s at 53 °C, 60 s at 72 °C; followed by 30 min of completing final extension at 60 °C. Mix 2: 5 min of initial denaturation at 95 °C; 18 cycles of 30 s at 94 °C, 90 s at 53 °C, 60 s at 72 °C; 11 cycles of 30 s at 94 °C, 90 s at 55 °C, 60 s at 72 °C; followed by 30 min of completing final extension at 60 °C. Mix 3: 5 min of initial denaturation at 95 °C; 14 cycles of 30 s at 94 °C, 90 s at 56 °C, 60 s at 72 °C; 11 cycles of 30 s at 94 °C, 90 s at 57 °C, 60 s at 72 °C; followed by 30 min of completing final extension at 60 °C. Mix 4: 5 min of initial denaturation at 95 °C; 24 cycles of 30 s at 94 °C, 90 s at 58 °C, 60 s at 72 °C; followed by 30 min of completing final extension at 60 °C. After amplification, PCR products of mix 1 and mix 4 were analysed separately while mix 2 and mix 3 were combined before loading onto the sequencer. Then, 1.5 µl of the PCR products of mix 1 and 4 and 3 µl of the combined PCR products of mix 2 and 3 were added to 13 µl formamide containing the GeneScan 500 LIZ Size Standard, heat denatured and resolved in POP4 polymer on an ABI 3100 Genetic Analyzer (all from Applied Biosystems, Darmstadt, Germany). Raw data were analysed and alleles assigned using GeneMapper 4.0. All eggs were fertilised and all eggs could be genotyped.

Parentage analysis was performed as described elsewhere[94]. Because all breeding females were known, both the probability of false exclusion of the social male and that of false inclusion of an extra-pair sire are low (~3 × $10^{-7}$ and ~$10^{-5}$, respectively). An egg was assigned to an extra-pair male when this male was the only candidate assigned with high probability and a maximum of 1 mismatch. For one clutch (5 eggs), paternity could not be assigned because the social father was unknown; for one egg from another clutch the quality of the genotyping allowed determination of embryo sex but not paternity. Among the 108 eggs whose paternity could be assigned, 13 (12%) were sired by extra-pair males.

**Protein extraction**. Albumen proteins were extracted in 4% SDS, 0.05 M TEAB buffer (14 ml extraction volume per 1 ml albumen) for 5 min at 95 °C under shaking. After 5 min of centrifugation at 4369 × $g$ the supernatant was collected and stored at −80 °C until analysis. Yolk samples (0.13–0.22 g) diluted with 200 µl ultra-pure water were delipidated prior to protein extraction by sequential addition of 1.6 ml methanol, 0.8 ml chloroform and 1.2 ml water. After each step the sample was homogenised and centrifuged for 1 min at 4369 × $g$. The upper water phase was removed, dried in a speed vac and dissolved in 1 ml sample buffer (first protein extract). The proteins were precipitated from the lower phase with 1.2 ml methanol and pelleted by centrifugation for 2 min at 4369 × $g$. The lipid phase was removed and frozen until further processing. The protein pellet was dissolved in 3 ml sample buffer containing 8 M urea, 2 M thiourea and 50 mM TEAB, and then mixed with the first protein extract. Dithiothreitol (DTT) was added to a final concentration of 5 mM to clarify the cloudy extract. Protein concentration was measured using a tryptophan assay against a BSA standard curve.

For each sample, 50 µg of proteins were diluted to 50 µl, reduced with 10 mM DTT, alkylated with 50 mM iodoacetamide and precipitated with 4 volumes of acetone overnight at −20 °C. The pellet was washed with 80% acetone, dried and dissolved in 0.1 M TEAB. Proteins were digested with 2 µg LysC for 4.5 h at 500 rpm, then with 0.5 µg trypsin overnight at 37 °C and 500 rpm. The resulting peptides were concentrated and desalted on C18 Stage Tips[95].

**Mass spectrometry analysis**. For analysis a Thermo EASY-nLC 1000 UHPLC system (Thermo Fisher Scientific, Bremen, Germany) was coupled online to a Q Exactive Orbitrap HF mass spectrometer with a nano-electrospray ion source (Thermo Fisher Scientific). The analytical column (15 cm long, 75 µm inner diameter) was packed in-house with ReproSil-Pur C18 AQ 1.9 µm reversed-phase resin (Dr. Maisch GmbH, Ammerbuch, Germany) in buffer A (0.5% formic acid). During online analysis, the analytical column was placed in a column heater (Sonation GmbH, Biberach, Germany) regulated to a temperature of 55 °C. Peptides were loaded onto the analytical column with buffer A at a back pressure of 980 bar (generally resulting in a flow rate of 700 nl/min) and separated with a linear gradient of 8–30% buffer B (80% acetonitrile and 0.5% formic acid) at a flow rate of 450 nl/min controlled by IntelliFlow technology over 24 min, (generally at a back pressure of around 500 bar). Online quality control was performed with SprayQc[96]. The mass spectrometry acquisition method was programmed with a data-dependent top 15 method, dynamically choosing the most abundant not yet sequenced precursor ions from the survey scans (300–1650 Th). The instrument itself was controlled using Tune 2.4 and Xcalibur 3.0. At a maximum ion inject time of 60 ms, the cycle time was ~1 s, sufficient for generating a median of 10 data points over the observed elution time of 20 s. Further settings were chosen according to their previously determined optimal values[97]. Sequencing was done with higher-energy collisional dissociation (HCD) fragmentation with a target value of 1e5 ions determined with predictive automatic gain control, for which the isolation of precursors was performed with a window of 1.4 Th. Survey scans were acquired at a resolution of 60,000 at $m/z$ 200 and the resolution for HCD spectra was set to 15,000 at $m/z$ 200. Normalised collision energy was set to 27 and the underfill ratio, specifying the minimum percentage of the target ion value likely to be reached at maximum fill time, was defined as 10%. This slightly elevated sequencing threshold ensured that, with the reduced complexity of samples, the fragmentation scans are of higher quality. Furthermore, the S-lens radio frequency level was set to 60, which gave optimal transmission of the $m/z$ region occupied by the peptides from our digest[97]. We excluded precursor ions with unassigned, single or five and higher charge states from fragmentation selection. Samples were measured in 1–3 technical replicates and were randomly assigned to analysis batches. The mass spectrometry proteomics data have been deposited to the ProteomeXchange Consortium via the PRIDE[98] partner repository with the dataset identifier PXD009822.

**Protein identification and quantitation**. The acquired mass spectrometry data were analysed with the MaxQuant proteomics data analysis workflow version 1.5.1.1[99]. The false discovery rate (FDR) cutoff was set to 1% for protein, peptide and peptide spectrum matches. Peptides were required to have a minimum length of 7 amino acids and a maximum mass of 4600 Da. MaxQuant was used to score fragmentation scans for identification based on a search with an initial allowed mass deviation of the precursor ion of a maximum of 4.5 ppm after time-dependent mass calibration. The allowed fragment mass deviation was 20 ppm. Fragmentation spectra were identified using the 6-frame translated blue tit transcriptome database[100] combined with 262 common contaminants by the integrated Andromeda search engine[101]. Enzyme specificity was set as C-terminal to arginine and lysine, also allowing cleavage before proline, and a maximum of two missed cleavages. Carbamidomethylation of cysteine was set as fixed modification and N-terminal protein acetylation and methionine oxidation as variable modifications.

We considered proteins correctly identified in a sample when we obtained positive intensity values from at least two LC-MS technical replicate runs. We considered proteins correctly quantitated when they satisfied the above identification criterion in at least half of the samples in at least one of the compared biological groups. For those proteins, missing values were replaced with random values extracted from a uniform distribution bounded by the minimum normalised intensity value and the 1% quantile of the intensity distribution for that sample and replicate combination. Intensity values were normalised by dividing them to the total MS signal for each sample and replicate. Normalised intensities were multiplied with the total amount of protein measured in the egg compartment and divided by the weight of the compartment to obtain an estimation of the concentration of the protein[102]. Although there is little known about the concentrations of individual proteins in passerine eggs, we were able to compare the concentrations obtained from this calculation to published values for three albumen proteins: ovotransferrin (6.7 mg/ml in this study; 26.8 ± 35.6 mg/ml in blue tits;[28] 3.63 ± 0.46 mg/ml in blue tits;[31] 4.34 ± 0.35 mg/ml in great tits;[31] 2.5–25.4 mg/ml in 9 species of larks *Alaudidae*;[103] 13.6–28.4 mg/ml in 8 species from 6 families[9]), avidin (2.7 μg/ml in this study; 1.1 ± 9 μg/ml in blue tits;[28] 0.23 ± 0.035 μg/ml in blue tits;[31] 0.32 ± 0.034 μg/ml in great tits;[31] 0.02–0.88 μg/ml in 8 species from 6 families;[9] 0.65 μg/ml yellow-legged gull *Larus michahellis*[23]) and lysozyme (223 μg/ml in this study; 107 μg/ml in blue tits;[104] 3.2–5910 μg/ml in 8 species from 6 families;[9] 15.7 μg/ml yellow-legged gull[23]). The match between our estimations and earlier results confirmed that these values can be used as proxy for protein concentration. Moreover, to obtain robust model estimates, protein concentration estimates were rank-transformed[105]. To account for the technical variation, the abundances of individual proteins in each sample were estimated from linear mixed-effect models with batch as random intercept and sample as a fixed factor.

Functional annotation was based on Gene Ontology and the MIPS Functional Catalogue database[42].

**Total lipids**. Total lipids were measured gravimetrically in the lipid phase obtained from the egg yolk delipidation. Aliquots of 100 μl of the lipid phase were transferred into pre-weighed 0.15 ml tin capsules (IVA Analysentechnik). The capsules were re-weighed after drying to constant mass and the lipid weight was obtained by subtracting the weight of the empty capsule. The measurements were performed in triplicate.

**Carotenoids**. Carotenoid identification: To ensure sufficient signal during liquid chromatography-tandem mass spectrometry (LC-MS/MS) analyses, a sample clean-up by saponification was required to eliminate co-eluting ion-suppressing lipids. For this purpose, 8 samples of the lipid phases (see Protein extraction) were pooled, saponified overnight with 30% (w/v) methanolic KOH and extracted from the saponified sample with hexane and diethyl ether. After washing each organic phase with water, the extracts were combined and evaporated to dryness, re-dissolved in 300 μl of a 1:1 mixture (v/v) of tert-butyl methyl ether (TBME) and methanol, and filtered (0.45 μm PTFE) into HPLC vials. For subsequent LC-MS/MS analyses, an Agilent 1100 series HPLC system (Agilent, Waldbronn, Germany) with ultraviolet/visible-photodiode array (UV/Vis-PDA) detector model G1315B was coupled online to a Bruker Esquire 3000+ (Bruker, Bremen, Germany) ion-trap mass spectrometer operated with an atmospheric pressure chemical ionisation (APCI) probe as described before[106]. HPLC separation was carried out according to ref. [107], except for reducing the flow rate from 1.8 to 1.2 ml/min, extending the linear separation gradient from 16 to 24 min and reducing column temperature from 40 to 35 °C for better compound separation (Supplementary Fig. 1a and b). Carotenoids were identified by comparison of UV/Vis spectra and mass spectra with literature data[108] and with those of authentic standards, i.e., β carotene, γ-carotene, lutein, β-zeacarotene and zeaxanthin (Supplementary Fig. 1c).

Carotenoid quantitation: An aliquot of 500 μl of the lipid phase (see Protein extraction) was evaporated to dryness. Carotenoid extraction was performed by sequential addition of 150 μl TBME and 150 μl methanol each followed by ultrasonication for 5 s. The extract was filtered through a 0.45 μm PTFE filter into an amber glass HPLC vial. Subsequent HPLC-PDA analyses were carried out on a Waters HPLC 2695 Separation Module coupled to a 2996 PDA detector, using the parameters described above. Carotenoids were quantitated at their respective UV/Vis absorption maxima using linear calibration curves of the above-mentioned standards. Unknown carotenoids were quantitated as lutein equivalents.

**Randomisation and statistical analysis**. All samples were processed in batches using a randomised block design. The samples were randomly reassigned to batches at each step of the protocol.

We used R version 3.4.3[109] for all analyses. Package lme4 was used to fit mixed-effects models with Gaussian error distribution[110]. Between- and within-clutch variation of egg weight and composition (Table 1) were computed from intercept-only, linear mixed-effect models with female identity as random intercept. Models on individual proteins used protein concentration as the dependent variable. Models on functional protein categories used as dependent variables either the combined concentrations or combined relative abundances (proportion of the total protein content of the egg compartment) of the proteins within a category, computed for the first-level FunCat terms. The sums were first computed for each replicate and then averaged per sample. Female identity was included as random intercept and laying order (the egg's position in the laying sequence) as random slope. Carotenoid concentrations were log transformed towards a normal distribution. Models on proteins and carotenoids included the MS, respectively the HPLC analysis batch as additional random intercept. Models with paternity as predictor are applied to first eggs of the clutches (13 extra-pair eggs vs. 24 within-pair eggs). Continuous (predictor or response) variables were standardised by centring and dividing by 2 SDs[111]. Estimates in all figures and tables are standardised effect sizes. To obtain robust model estimates from data containing extreme values, protein concentrations were rank-transformed[105]. Within-model $p$ value correction was performed using single-step error correction as implemented in the multcomp package[112]. We also applied a multiple testing correction (FDR)[113] when at least 5 models were run. We tested three a priori chosen interactions between predictors: (1) female age × laying order, (2) female age × clutch size and (3) embryo sex × laying order. All tests are two-tailed unless otherwise specified.

The slopes of egg-weight variation with the laying order did not differ when estimated based on all eggs sampled from a clutch or solely based on eggs 1 and 9 (linear mixed-effect model with egg weight as dependent variable and the interaction between laying order and sampling type (all eggs vs. eggs 1 and 9): $z = -0.09$, $p = 0.83$).

We used affinity propagation clustering (AP clustering)[114] implemented in the R package apcluster (v.1.4.4)[115] for unsupervised detection of protein expression patterns and of egg groups with homogenous composition. The algorithm determines the optimal number of clusters based on a similarity matrix (in this case pairwise Pearson's correlation coefficients between log-transformed concentrations of all proteins). The exemplar preference threshold parameter[114] (apcluster function argument $q$)[115] was set to 0 to obtain a conservative estimate of the number of clusters. Because the deposition of yolk and albumen proteins is spatially and temporally segregated in the body of the breeding female (ovary and oviduct, respectively) and under the influence of different physiological and environmental factors, the albumen and yolk proteins were clustered separately to allow for the emergence of divergent patterns.

Protein-cluster enrichment in first-level FunCat terms was determined using a Monte-Carlo simulation. The proteins were randomly assigned to the clusters and the frequencies of the 21 FunCat terms were computed for each cluster. The cycle was repeated 10,000 times and a distribution of frequencies was obtained for each cluster and FunCat term pair. A cluster was considered enriched in a protein function if the observed frequency of the corresponding FunCat term exceeded the 95% quantile of the simulated frequency distribution obtained for that cluster and FunCat term pair. Protein concentrations were combined per cluster and modelled using the same procedure described for the functional protein categories.

AP clustering was also used as a dimensionality-reduction method for the protein and the carotenoid datasets[116]. The clusters of proteins (7 albumen clusters, 9 yolk clusters) were used as input variables together with the concentration of yolk lipids and the egg weight for the clustering of eggs. Because AP clustering only returned one carotenoid cluster, the total carotenoid concentration was used as input variable in this analysis. We used functional chi-square tests[117] to determine whether any of the following variables significantly predicted the egg clusters: clutch size, lay date (date of the first egg of the clutch), laying order, female age, embryo sex and paternity (within-pair vs. extra-pair). Non-binary variables were categorised into three groups based on equal intervals (lay date: 4-day intervals, i.e., early-, middle- and late-laid first eggs; laying order: early = eggs 1 and 3, middle = eggs 5 and 7, late = eggs 9, 11 and 13; clutch size: small = 8–10 eggs, average = 11–12 eggs, large = 13–15 eggs).

**Code availability**. Computer code used for figure production can be found at https://osf.io/nqw6v/.

**Data availability**
The mass spectrometry proteomics data have been deposited to the ProteomeXchange Consortium via the PRIDE[98] partner repository with the dataset identifier PXD009822. The datasets generated during the current study are available at https://osf.io/nqw6v/.

## Acknowledgements

We thank Matthias Mann for hosting the mass spectrometry analysis in his lab. We thank Andrea Wittenzellner, Agnes Türk and Alexander Girg in the B.K. group, and Gaby Sowa, Igor Paron, and Korbinian Mayr in the group of Matthias Mann for technical support. We thank Jakob Müller and Heiner Kuhl for making blue tit sequence data available ahead of publication. We are grateful to Jarrod Hadfield for providing valuable comments on the manuscript.

## Author contributions
Conceived of the study: C.M.V. and B.K.; developed the methods: C.M.V., R.A.S., R.M.S., K.T., M.V., D.W., R.C.; performed experiments and produced the data: C.M.V., R.A.S., R.M.S., K.T.; analysed the data: M.V. and C.M.V.; wrote the paper: C.M.V. and B.K. with input from all authors.

## Additional information
**Supplementary information** accompanies this paper at https://doi.org/10.1038/s42003-018-0247-8.

**Competing interests:** During the study, one of the authors (R.M.S.) left the University of Hohenheim to take up a position at a manufacturer of carotenoid supplements (DSM Nutritional Products, Kaiseraugst, Switzerland). The remaining authors declare no competing interests.

**Reprints and permission** information is available online at http://npg.nature.com/reprintsandpermissions/

**Publisher's note:** Springer Nature remains neutral with regard to jurisdictional claims in published maps and institutional affiliations.